\documentclass[journal]{IEEEtran}
\usepackage[latin9]{inputenc}
\usepackage{amsmath}
\usepackage{amssymb}
\usepackage{graphicx}

\makeatletter

\usepackage{epsfig}\usepackage{amsfonts}\usepackage{array}\usepackage{dcolumn}\usepackage{psfrag}\usepackage{amsfonts}\usepackage{accents}\usepackage{booktabs}
\usepackage{cite}\usepackage{subfigure}\usepackage[all]{xy}
\usepackage{dsfont}\usepackage{url}\usepackage{xcolor}\usepackage[nolist]{acronym}

\makeatother

\begin{document}


\title{Interference Spins: Scheduling of Multiple Interfering Two-Way Wireless
Links}

\author{
Petar~Popovski,~\IEEEmembership{Senior Member,~IEEE}, Osvaldo~Simeone,~\IEEEmembership{Senior Member,~IEEE},\\ Jimmy J. Nielsen,~\IEEEmembership{Member,~IEEE} and \v Cedomir Stefanovi\'c,~\IEEEmembership{Member,~IEEE}
\thanks{P. Popovski, J. J. Nielsen and \v C. Stefanovic\' c are with the Department of Electronic Systems, Aalborg University, 9220 Aalborg, Denmark (e-mail: \{petarp, jjn, cs\}@es.aau.dk).}
\thanks{Osvaldo Simeone is with CWCSPR, New Jersey Institute of Technology, Newark, NJ 07102 USA (e-mail: osvaldo.simeone@njit.edu).}
}

\maketitle
\begin{abstract}
Two-way is a dominant mode of communication in wireless systems.
Departing from the tradition to optimize each transmission direction separately,
recent work has demonstrated that, for time-division duplex (TDD)
systems, optimizing the schedule of the two transmission directions
depending on traffic load and interference condition leads to
performance gains. In this letter, a general network of multiple interfering
two-way links is studied under the assumption of a balanced load in
the two directions for each link. Using the notion of \emph{interference spin}, we introduce an algebraic framework
for the optimization of two-way scheduling, along with an efficient optimization
algorithm that is based on the pruning of a properly defined
topology graph and dynamic programming. Numerical results demonstrate
multi-fold rate gains with respect to baseline solutions, especially
for worst-case (5\%-ile) rates.

\end{abstract}

\section{Introduction}

Two-way is a dominant mode of communication in wireless systems, such
as in uplink (UL)/downlink (DL) cellular communication. Although the
two-way channel is the first known multi-user channel treated in information
theory \cite{Shannon}, the design of two-way links has traditionally
been tackled by separating the problem into independent one-way DL and UL problems, respectively,
that deal with rate adaptation, scheduling, etc.

Nevertheless, recent works \cite{Yu2012,Shen2013,Zhu2013,Vent2014,Dow2013,Bamby2014}
have demonstrated that the presence of interference motivates the joint consideration of UL and DL, since the interference caused by a link on other links
is different depending on the direction in which the link is active,
i.e., on which node acts as the transmitter. Specifically, the referenced works have studied dynamic Time Division
Duplex (TDD) for small-cell wireless systems.
With dynamic TDD, each slot, of possibly different size in the frequency-time
plane, can be assigned to either the UL or DL direction, depending
on traffic load and interference conditions. References \cite{Yu2012}
and \cite{Shen2013} outline the challenge of managing UL-DL cross-interference
in the presence of dynamic TDD. The works \cite{Dow2013,Bamby2014}
instead propose centralized and decentralized algorithms that optimize
the switching time between UL and DL. Finally, \cite{Zhu2013,Vent2014}
put forth heuristic solutions for the optimization of UL-DL slot allocation
for fixed switching times.

\begin{figure}[t]
\centering \includegraphics{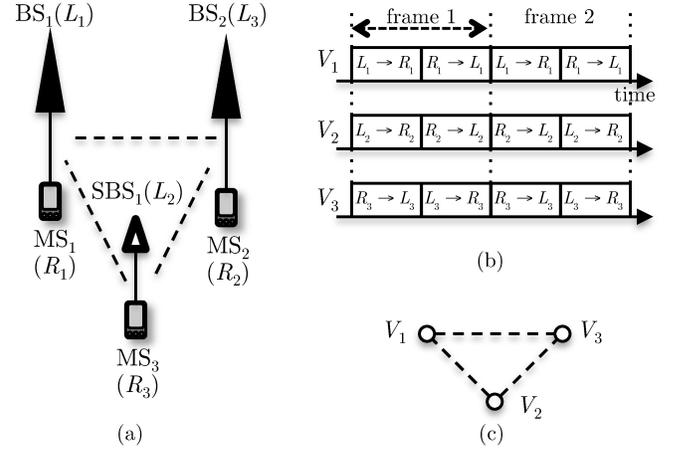} \protect\protect\protect\protect\caption{(a) Example scenario with two macro-cells and one small cell, with
direct channels (full lines) and interfering channels (dashed lines).
(b) Example spin configuration
in two subsequent frames. (c) Topology graph $\mathcal{G}$.}

 \label{fig:SpinExample1}
\end{figure}

In this letter we consider a generic collection of two-way links,
hence including, e.g., cellular and/or device-to-device (D2D) links, where the traffic load is balanced between the two transmission directions: a slot assigned to one transmission direction is always followed
by one allocated to the opposite direction. This assumption is per definition suitable for
applications using interactive communication. An example
setup is given in Fig.~\ref{fig:SpinExample1}(a), that consists
of three interfering two-way links operating over two-slot frames,
with one slot allocated to each direction as per Fig.~\ref{fig:SpinExample1}(b).
We refer to the binary variable that determines the link direction in the first slot (the opposite to the one in the second slot) as
the \emph{interference spin} of a link (see precise definition in
the next section). This allows us to propose an \emph{algebraic framework}
for the optimization of two-way scheduling, e.g., UL-DL slot allocation. The optimization consists of selecting the interference spins
of all $M$ links in order to maximize a general sum-utility of the signal-to-noise-plus-interference
ratios (SINRs) of the two-way links. The optimal solution has a complexity
that scales as $2^{M}$. Besides the algebraic framework, our second contribution is
a suboptimal, but efficient,
algorithm that leverages the representation of the $M$ interfering
two-way links as a graph of $M$ nodes. The proposed algorithm is
based on pruning this topology graph according to a specific criterion
and applying dynamic programming. The algorithm finds a spin configuration
that is seen via numerical results to exhibit substantial performance
gains over the baseline scheme in which the spins are not optimized.

\section{System Model and Problem Definition\label{sec:System-Model-and}}

We consider a set $\mathcal{V}$ of $M$ synchronous two-way links,
where the $l-$th link $V_{l}$ consists of two half-duplex nodes
$L_{l}$ and $R_{l}$. The labeling of the two nodes in a link $V_{l}$
as left, $L_{l}$, and right, $R_{l}$, is arbitrary, see Fig.~\ref{fig:SpinExample1}.
Each link $V_{l}$ uses TDD, and two-way communication takes place
in a frame consisting of two slots. The first slot is used in one
direction, either $L_{l}\rightarrow R_{l}$ or $R_{l}\rightarrow L_{l}$,
and the successive slot is used in the opposite direction. Related
to this, we define \emph{interference spin} or, for short, \emph{spin}
of a link in a given frame: the $l-$th link $V_l$ is said to have a \emph{left
spin} or $0-$spin if $L_{l}$ transmits in the odd slot and $R_{l}$
transmits in the even slot; otherwise the link has a \emph{right spin}
or $1-$spin in that frame. In the example in Fig.~\ref{fig:SpinExample1}(b),
the link $V_{2}$ has a left spin in the first frame and a right spin
in the second frame. The spin of the $l-$th link is denoted by $s_{l}\in\{0,1\}$.

\begin{figure}[t]
\centering \includegraphics{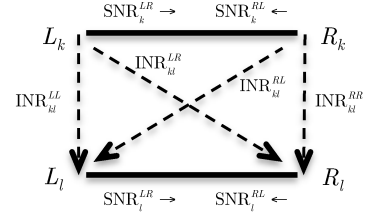} \protect\protect\caption{Two interfering links $V_{k}:\textrm{ }L_{k}-R_{k}$ and $V_{l}:\textrm{ }L_{l}-R_{l}$.
The INRs are given only for the interference from $V_{l}$ to $V_{k}$
for clarify of illustration.}

 \label{fig:TwoPairs}
\end{figure}

All devices use the same spectrum in a TDD manner, leading to interference among the
concurrent link transmissions. Each device is backlogged with data,
such that there is transmission in each slot and there is a continuous
stream of slots. All links are slot-/frame-synchronous. Considering
two links, say $V_{k}: L_{k}-R_{k}$ and $V_{l}: L_{l}-R_{l}$,
we distinguish the following channel gains, see Fig.~\ref{fig:TwoPairs}:
(a) \emph{Direct channels}: $L_{k}\rightarrow R_{k}$, $R_{k}\rightarrow L_{k}$,
$L_{l}\rightarrow R_{l}$ and $R_{l}\rightarrow L_{l}$, whose signal-to-noise
ratios (SNRs) are defined as $\mathrm{SNR}_{k}^{LR}$, $\mathrm{SNR}_{k}^{RL}$,
$\mathrm{SNR}_{l}^{LR}$ and $\mathrm{SNR}_{l}^{RL}$, respectively;
(b) \emph{Interfering channels}: $L_{k}\rightarrow R_{l},$ $R_{k}\rightarrow L_{l}$,
$L_{k}\rightarrow L_{l}$ and $R_{k}\rightarrow R_{l}$, whose interference-to-noise
ratios (INRs) are defined as $\mathrm{INR}_{kl}^{LR}$, $\mathrm{INR}_{kl}^{RL}$,
$\mathrm{INR}_{kl}^{LL}$ and $\mathrm{INR}_{kl}^{RR}$, respectively.
The interference from $l$ to $k$, not depicted in Fig.~\ref{fig:TwoPairs},
is represented in a similar manner. Even with perfect channel reciprocity,
we generally have $\mathrm{SNR}_{k}^{LR}\neq\mathrm{SNR}_{k}^{RL}$
and $\mathrm{INR}_{kl}^{LR}\neq$$\mathrm{INR}_{lk}^{RL}$, since
the nodes at the two ends of a link may use different powers. Some
of these quantities can be zero: for instance, links may not interfere
with each other, e.g. due to obstacles and yield $\mathrm{INR}_{lk}^{LR}=$$\mathrm{INR}_{kl}^{RL}=0$.



In order to capture the interference-related features of the network,
we define the undirected \emph{topology graph} $\mathcal{G}=(\mathcal{V},\mathcal{E})$,
where the set of vertices $\mathcal{V}$ represents the \emph{links}
$V_{l}$, as defined above, and an edge exists in the edge set $\mathcal{E}$
for a pair of links $V_{k}$ and $V_{l}$ in $\mathcal{L}$ if and
only if at least one of the interfering powers $\mathrm{INR}_{kl}^{LR}$,
$\mathrm{INR}_{kl}^{RL}$, $\mathrm{INR}_{kl}^{LL}$, $\mathrm{INR}_{kl}^{RR}$,
$\mathrm{INR}_{lk}^{LR}$, $\mathrm{INR}_{lk}^{RL}$, $\mathrm{INR}_{lk}^{LL}$
and $\mathrm{INR}_{lk}^{RR}$ is non-zero. Therefore, an edge $kl\in\mathcal{E}$
indicates that two links $V_{k}$ and $V_{l}$ interfere at least
in one direction. Note that we identify edges via the indices of the
connected links. The topology graph for the example in Fig.~\ref{fig:SpinExample1}(a)
is depicted in Fig.~\ref{fig:SpinExample1}(c). Without loss of generality,
the graph is assumed to be \emph{connected} \cite{Kruskal}, since,
otherwise, one could consider the different connected components separately.

Denoting by $\oplus$ the XOR operation, the interference between
two links $V_k$ and $V_l$ is fully specified by the \emph{relative spin
$r_{kl}$}:
\begin{equation}
r_{kl}=r_{lk}=s_{k}\oplus s_{l}.\label{eq:DefinitionRspin}
\end{equation}
The SINRs for any link $V_l$ in the two directions $L_{l}\rightarrow R_{l}$
and $R_{l}\rightarrow L_{l}$ can be written as a function solely
of the relative spins of the interfering links, i.e., of $r_{kl}$
with $kl\in\mathcal{E}$. Specifically, the SINR for the direction
$L_{l}\rightarrow R_{l}$ is given as
\begin{equation}
\mathrm{SINR}_{l}^{LR}(\mathbf{r}_{l})=\frac{\mathrm{SNR}_{l}^{LR}}{1+\sum_{k:\textrm{ \ensuremath{kl\in\mathcal{E}}}}[(1-r_{kl})\mathrm{INR}_{kl}^{LR}+r_{kl}\mathrm{INR}_{kl}^{RR}]}\label{eq:SINRlr}
\end{equation}
and for the direction $R_{l}\rightarrow L_{l}$ we have
\begin{equation}
\mathrm{SINR}_{l}^{RL}(\mathbf{r}_{l})=\frac{\mathrm{SNR}_{l}^{RL}}{1+\sum_{k:\textrm{ \ensuremath{kl\in\mathcal{E}}}}[(1-r_{kl})\mathrm{INR}_{kl}^{RL}+r_{kl}\mathrm{INR}_{kl}^{LL}]},\label{eq:SINRrl}
\end{equation}
where $\mathbf{r}_{l}$ is the vector of spins $r_{kl}$, $kl\in\mathcal{E}$,
for the link $V_l$. By rewriting the denominator of (\ref{eq:SINRlr})
as $1+\sum_{k:\textrm{ \ensuremath{kl\in\mathcal{E}}}}[\mathrm{INR}_{kl}^{LR}+r_{kl}(\mathrm{INR}_{kl}^{RR}-\mathrm{INR}_{kl}^{LR})]$,
we see that, if $\mathrm{INR}_{kl}^{RR}=\mathrm{INR}_{kl}^{LR}$,
the relative spin $r_{kl}$ does not affect $\mathrm{SINR}_{l}^{LR}(\mathbf{r}_{l})$.
Similarly, if $\mathrm{INR}_{kl}^{LL}=\mathrm{INR}_{kl}^{RL}$, then
$r_{kl}$ does not affect $\mathrm{SINR}_{l}^{RL}(\mathbf{r}_{l})$.

Given a topology graph $\mathcal{G}$, definition (\ref{eq:DefinitionRspin})
implies that the vector $\mathbf{r}=[\mathbf{r}_{1}\cdots\mathbf{r}_{M}]$
of all relative spins satisfies the following properties: \textbf{(C1)}
\emph{Symmetry}: $r_{kl}=r_{lk}$; \textbf{(C2)} \emph{Parity check
on cycles}: For any cycle in the topology graph $l_{1}l_{2},\textrm{ }l_{2}l_{3}...,\textrm{ }l_{N}l_{1}\in\mathcal{E}$,
we have the parity check equation
\begin{equation}
r_{l_{1}l_{2}}\oplus r_{l_{2}l_{3}}\oplus...\oplus r_{l_{N-1}l_{N}}\oplus r_{l_{N}l_{1}}=0.\label{eq:cycle}
\end{equation}
It can be easily shown that \textbf{C1} and \textbf{C2} are also sufficient
to guarantee the existence of a vector of spins that satisfies (\ref{eq:DefinitionRspin})
and hence they characterize the set of all relative spin vectors.
A simple consequence of \textbf{C1} and \textbf{C2} is that the specification
of the relative spins on the edges of any spanning tree \cite{Kruskal}
on the topology graph $\mathcal{G}$ is sufficient to obtain the entire
vector $\mathbf{r}$. Every edge $kl$ either belongs to the tree
and hence $r_{kl}$ is fixed, or, if not, then edge $kl$ forms a
unique cycle with a subset of the edges of the tree and the relative
spin follows from (\ref{eq:cycle}).


We are interested in finding the spin configuration that maximizes
a sum-utility function of the SINRs across all links. The problem
of interest is hence formulated as
\begin{align}
\textrm{\ensuremath{\ensuremath{\underset{\mathbf{r}}{\textrm{max}}}}} & \sum_{l=1}^{M}U_{l}(\mathrm{SINR}_{l}^{LR}(\mathbf{r}_{l}),\mathrm{SINR}_{l}^{RL}(\mathbf{r}_{l})) \quad \textrm{s.t. \textbf{C1} and \textbf{C2}}\label{eq:opt problem}
\end{align}
where the maximization over the relative spins is subject to the constraints
\textbf{C1} and \textbf{C2, }and we have fixed non-negative utility
functions $U_{l}(\cdot,\cdot)$ for each link $V_{l}$. The utility
function $U_{l}(\cdot,\cdot)$ is assumed to be non-decreasing in
the arguments and will be written as $U_{l}(\mathbf{r}_{l})$ in order
to simplify the notation. An example is the two-way sum-rate, which
is given as $U_{l}(\mathbf{r}_{l})=\log(1+\mathrm{SINR}_{l}^{LR}(\mathbf{r}_{l}))+\log(1+\mathrm{SINR}_{l}^{RL}(\mathbf{r}_{l}))$.
In principle, a link can change the spin per frame; however, in Sec. \ref{sec:Numerical-Results}, we will treat
the case in which a link spin is determined based on SINR variables that include long-term fading and update the  the scheduling decisions on a large time scale.

\section{Optimizing the Interference Spins}

The maximization in (\ref{eq:opt problem}), in principle, can be
carried out by exploring all possible configurations of relative spins.
Due to the constraint (\ref{eq:cycle}), the complexity of exhaustive
search scales as $2^{M-1}$, becoming impractical as $M$ increases.
Here we propose a suboptimal, but computationally more efficient optimization
strategy that consists of two steps: 1) Construction of a \emph{maximum
relative-interference} \emph{spanning tree} $\mathcal{T}$ over the
graph $\mathcal{G}$; 2) Dynamic programming spin optimization on
$\mathcal{T}$.

\subsection*{1) Maximum Relative-Interference Spanning Tree $\mathcal{T}$}

For each edge $kl$ in $\mathcal{G}$, a weight $w_{kl}$ is assigned
as
\begin{align}
w_{kl}=\max(|\mathrm{INR}_{kl}^{RR}-\mathrm{INR}_{kl}^{LR}|,|\mathrm{INR}_{kl}^{LL}-\mathrm{INR}_{kl}^{RL}|,\nonumber \\
|\mathrm{INR}_{lk}^{RR}-\mathrm{INR}_{lk}^{LR}|,|\mathrm{INR}_{lk}^{LL}-\mathrm{INR}_{lk}^{RL}|).\label{eq:relative interference}
\end{align}
This weight evaluates the maximum change in interference powers that
is affected by the selection of the spin $r_{kl}$. A maximum spanning
tree $\mathcal{T}$ is then constructed with respect to the weights
in (\ref{eq:relative interference}), retaining only the edges that
have the largest contributions to the relative interference powers
and pruning the remaining edges; e.g., for the example in Fig.~\ref{fig:SpinExample1}(c), one of the three edges will be removed based on the criterion (\ref{eq:relative interference}), thus obtaining $\mathcal{T}$ with two edges. We refer to $\mathcal{T}$ as the maximum relative-interference spanning tree. The tree $\mathcal{T}$ can be constructed
in a centralized or distributed way \cite{Kruskal}\cite{Gallager1983},
but the implementation details are out of the scope of this paper.
Finally, a root vertex is arbitrarily selected so as to make $\mathcal{T}$
a rooted tree. Note that each vertex $V_{l}$ in $\mathcal{T}$ has
a single parent vertex $P_{l}$ (except for the root vertex, which has
no parent), i.e., $P_{l}$ is the only vertex in $\mathcal{T}$ along
the unique path to the root. $V_{l}$ can be the parent for multiple \emph{children} vertices $\mathcal{C}_{l}$.

\subsection*{2) Dynamic Programming}

Having constructed the spanning tree $\mathcal{T}$, we now proceed
to optimize only the relative spins corresponding to the edges $kl\in\mathcal{T}$
via dynamic programming. As discussed in Sec. \ref{sec:System-Model-and},
the spins for all the remaining edges can then be immediately calculated
via (\ref{eq:cycle}). In order to allow the optimization to be limited
only to the edges in $\mathcal{T}$, we approximate the SINRs in \eqref{eq:SINRlr}
and \eqref{eq:SINRrl} so that they only depend on the relative spins
for the edges in $\mathcal{T}$. To this end, denote by $\mathbf{r}_{\mathcal{C}_{l}l}$
the vector of the relative spins for the edges connecting the child
nodes $\mathcal{C}_{l}$ with $V_{l}$ and, similarly, by $r_{lP_{l}}$
the relative spin between $V_{l}$ and its parent $P_{l}$. The SINR
\eqref{eq:SINRlr} for link $V_{l}$ is approximated as
\begin{align}
 & \hat{\mathrm{SINR}}{}_{l}^{LR}(\mathbf{r}_{\mathcal{C}_{l}l},r_{lP_{l}})=\frac{\mathrm{SNR}_{l}^{LR}}{1+\sum_{k:\textrm{ \ensuremath{kl\in\mathcal{E}}}}\mathrm{\hat{INR}}_{kl}},\text{ where}\label{eq:approximate SINR}\\
 & \mathrm{\hat{INR}}_{kl}=\left\{ \begin{array}{ll}
(1-r_{kl})\mathrm{INR}_{kl}^{LR}+r_{kl}\mathrm{INR}_{kl}^{RR} & k\in\mathrm{\mathcal{C}}_{l}\cup\{P_{l}\},\\
\mathrm{INR}_{kl}^{LR}+(\mathrm{INR}_{kl}^{LR}+\mathrm{INR}_{kl}^{RR})/2 & \text{else}.
\end{array}\right.
\end{align}
In other words, the interference contribution for the edges $kl\in\mathcal{E}\setminus\mathcal{T}$
that do not belong to the tree is approximated with the average of
the interference powers that would be observed if $r_{kl}=0$ or $r_{kl}=1$.
This is justified by the fact that, in light of the choice of the
weights (\ref{eq:relative interference}), it is expected that the
two values $\mathrm{INR}_{kl}^{LR}$ and $\mathrm{INR}_{kl}^{RR}$
are similar. The same approach is used to approximate \eqref{eq:SINRrl},
leading to the approximation $\hat{\mathrm{SINR}}{}_{l}^{RL}(\mathbf{r}_{\mathcal{C}_{l}l},r_{lP_{l}})$.
For ease of notation, we write $\hat{U}_{l}(\mathbf{r}_{\mathcal{C}_{l}l},r_{lP_{l}})=U_{l}(\hat{\mathrm{SINR}}{}_{l}^{LR}(\mathbf{r}_{\mathcal{C}_{l}l},r_{lP_{l}}),\hat{\mathrm{SINR}}{}_{l}^{RL}(\mathbf{r}_{\mathcal{C}_{l}l},r_{lP_{l}}))$.

We now aim at optimizing problem (\ref{eq:opt problem}) with utilities
$\hat{U}_{l}$ in lieu of $U_{l}$ for all $l\in\mathcal{V}$. The
proposed dynamic programming solution starts from the leaf vertices and
proceeds according to the (partial) order defined by the tree until
the root is reached. In particular, each vertex $V_{l}$ calculates
the message $\boldsymbol{\mu}_{l}=(\mu_{l}^{0},\mu_{l}^{1})$ for
its parent vertex $P_{l}$, where
\begin{align}
\mu_{l}^{i}=\max_{\mathbf{r}_{\mathcal{C}_{l}l}}\left(\hat{U}_{l}\left(\mathbf{r}_{\mathcal{C}_{l}l},r_{lP_{l}}=i\right)+\sum_{k\in\mathcal{C}_{l}}\mu_{k}^{r_{kl}}\right),\label{eq:message}
\end{align}
for $i=0,1$. The maximization in (\ref{eq:message}) is over the
relative spins corresponding to edges stemming from the child vertices
$\mathcal{C}_{l}$. We denote a solution of the problem (\ref{eq:message})
as $\mathbf{\bar{r}}_{\mathcal{C}_{l}l}^{i}$ for $i=0,1$. Note that
the leaves have no child vertices, and hence, for every leaf $V_{l}$,
the message is calculated as $\mu_{l}^{i}=\hat{U}_{l}(r_{lP_{l}}=i)$
for $i=0,1$. Instead, the root vertex $t$, which has no parent, solves
the problem $\max_{\mathbf{r}_{\mathcal{C}_{t}t}}(\hat{U}_{t}(\mathbf{r}_{\mathcal{C}_{t}t})+\sum_{k\in\mathcal{C}_{t}}\mu_{k}^{r_{kt}}$)
and obtains an optimal solution $\bar{\mathbf{r}}_{\mathcal{C}_{t}t}$.
A complete solution $\bar{\mathbf{r}}$ is
finally obtained by backpropagation: starting with the children of
the root vertex, each child vertex $k\in\mathcal{C}_{l}$ of a vertex $V_{l}$
selects the solution $\mathbf{\bar{r}}_{\mathcal{C}_{l}l}^{\bar{r}_{kl}}$,
until the process reaches the leaves.

The complexity of the maximum spanning tree construction scales as
$G\log M$ \cite{Kruskal}\cite{Gallager1983}, where $G\leq M(M-1)/2$
and $M$ are the number of edges and vertices in $\mathcal{G}$, respectively.
The order of complexity of the relative-spin optimization is $M\cdot2^{D}$,
where $D$ is the maximum number of children of a vertex in $\mathcal{T}$.
While in the worst case, $D=M-1$, for a typical topology graph $\mathcal{G}$,
we found $D$ to be much smaller than $M$, leading to significant
complexity saving with respect to exhaustive search, see Section~\ref{sec:Numerical-Results}.

\section{Numerical Results and Concluding Remarks\label{sec:Numerical-Results} }

For performance evaluation, a small-cell set-up is considered over
a $100\text{m}\times100\text{m}$ area. $M$ links, or equivalently $2M$ nodes, are generated as follows: $M$ transceivers are placed uniformly
in this area and chosen equiprobably to be either an $L-$ or $R-$
node; then, the opposite node of the link ($R$ or $L$, respectively)
is placed at a uniformly selected random angle with distance $d$. Note that the opposite node may lie outside the area at hand.
Two types of links are considered: (\emph{i}) \emph{symmetric D2D
links}, with $d=d_{s}=10$m; (\emph{ii}) \emph{asymmetric femtocell
links} with $d=d_{a}=50$m. These links differ as explained below. The long-term SNRs and INRs that are used
by the algorithm to determine the configuration of relative spins
account only for large-scale fading and path loss. The long-term SNR
parameters for a symmetric link $V_{l}$ are given as $\mathrm{SNR}_{l}^{RL}=\mathrm{SNR}_{l}^{LR}=\mathrm{SNR}_{s}\beta_{l}$. The transmit power is only adapted to the path loss in order to get $\mathrm{SNR}_{s}=20$dB, while
$\beta_{l}$ represent independent log-normal shadowing
with standard deviation $8$dB. For an asymmetric link $V_{l}$, the
SNRs are given as $\mathrm{SNR}_{l}^{LR}=\mathrm{SNR}_{a}^{LR}\beta_{l}$
and $\mathrm{SNR}_{l}^{RL}=\mathrm{SNR}_{a}^{RL}\beta_{l}$ with $\mathrm{SNR}_{a}^{LR}=20$dB
and $\mathrm{SNR}_{a}^{RL}=10$dB due to the fact that the power of
the femto-base station ($L$) is different from the power of the mobile
user ($R$). The long-term interference power caused by the node $X_{l}$
to the node $Y_{k}$, where $X,Y\in\{L,R\}$, is given as $\mathrm{INR}_{lk}^{XY}=\mathrm{SNR}_{z}^{X\bar{X}}(d_{z}/d_{lk}^{XY})^{\eta}\beta_{lk}^{XY}$,
where $z=s$ if $V_{l}$ is symmetric and $z=a$ otherwise, $X\bar{X}\in\{LR,RL\}$,
$\eta=4$ is the path loss exponent, $d_{lk}^{XY}$ is the distance
between node $X_{l}$ and $Y_{k}$, and $\beta_{lk}^{XY}$ accounts
for log-normal shadowing with standard deviation $8$dB. Note that
$\beta_{lk}^{XY}=\beta_{kl}^{YX}$.

For a fixed spatial configuration and large-scale fading, the spins
are optimized with the proportional fairness utility across the links,
so that $U_{l}(\mathbf{r}_{l})=\log(\log(1+\mathrm{SINR}_{l}^{LR}(\mathbf{r}_{l}))+\log(1+\mathrm{SINR}_{l}^{RL}(\mathbf{r}_{l}))).$
The proposed algorithm is referred to as MST-DP (Maximum Spanning
Tree-Dynamic Programming). One reference, when computationally feasible,
is the the optimal performance based on exhaustive search; the other
is the average performance with uniform random spins. For a fixed spatial configuration, large-scale fading and
relative spins, we evaluate for each link $V_{l}$ the instantaneous
two-way sum-rate $\log(1+\mathrm{sinr}_{l}^{LR}(\mathbf{r}_{l}))+\log(1+\mathrm{sinr}_{l}^{RL}(\mathbf{r}_{l}))$,
where the instantaneous SINRs $\mathrm{sinr}_{l}^{LR}(\mathbf{r}_{l})$
and $\mathrm{sinr}_{l}^{RL}(\mathbf{r}_{l})$ are calculated via (\ref{eq:SINRlr})
and (\ref{eq:SINRrl}) by multiplying the corresponding long-term
SNR and INR variables by unit-power Rayleigh variables, independently drawn for each par of nodes in each frame.

\begin{figure}[t]
\centering \includegraphics[width=0.5\textwidth]{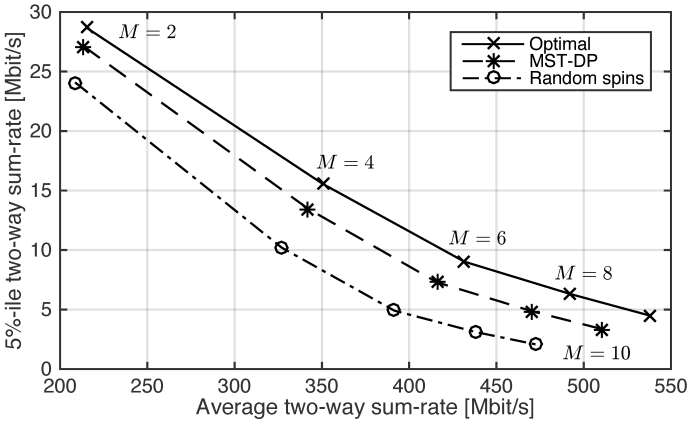}
\protect\protect\protect\protect\caption{5\%-ile two-way sum-rate (for a bandwidth of $10$MHz) versus the
average two-way sum-rate for different number of links $M$.}

\label{fig:results_sumVs5prct_fair} 

\end{figure}

In Fig. \ref{fig:results_sumVs5prct_fair}, we plot the 5\%-ile of
the two-way sum-rate, which is a common measure of worst-case rate,
against the average two-way sum-rate, for different number $M$ of
symmetric links and a bandwidth of $10$MHz. Note that having more
links $M$ enables the spatial reuse, and hence the sum-rate, to be
improved, but, at the same time, it creates more interference that
reduces the worst-case rates. It is seen that spin optimization can
significantly outperform random-spin assignment, especially in terms
of 5\%-ile rate, which, as an example, is nearly doubled for $M=10$.

\begin{figure}[t]
\centering \includegraphics[width=0.5\textwidth]{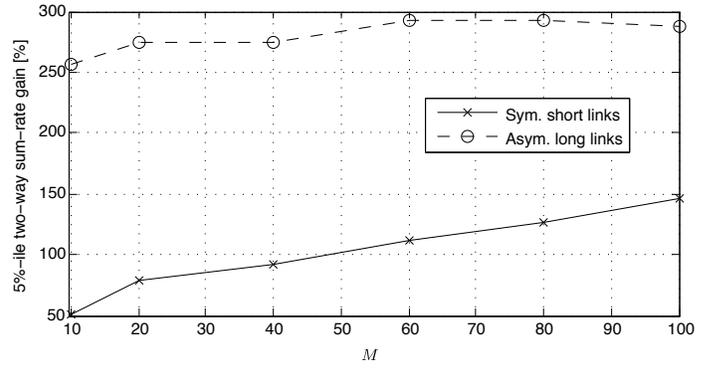}
\protect\protect\protect\protect\caption{Gain in $5\%$-ile two-way sum-rate for the proposed MST-DP algorithm
with respect to a random spin selection versus the number of links.}

\label{fig:Results_GainSymmetricAsymmetricVaryM}

\end{figure}

Fig.~\ref{fig:Results_GainSymmetricAsymmetricVaryM} elaborates further
on the 5\%-ile performance gains attained by optimized spin assignment
with respect to random-spin selection as a function of $M$. Note that the high number of links is justified by the ultra-dense wireless scenarios envisioned in the upcoming 5G wireless systems~\cite{Vent2014}.
Two curves
are shown, one for all symmetric and one for all asymmetric links,
respectively. For computational feasibility, only the MST-DP algorithm
is considered. It is seen that, while the performance gains are substantial
in both scenarios, optimizing the interference spins is particularly
advantageous for asymmetric links. This is expected, since controlling
the interference caused by the nodes that transmit with larger power
has a more pronounced impact on the worst-case two-way sum-rate. Finally, the MST-DP algorithm, while inferior to the exhaustive search,
still provides significant gains over random spins. We observe that,
in terms of complexity, even with $M=100$ nodes, the value of $D$
was found to be always $D \leq 9$ and to be, on average less than $6$.

The main conclusion is that the degree of freedom offered by the interference
spins leads to remarkable throughput
gains, especially for the worst-case (5\%-ile) rates.

\end{document}